\documentclass[prl,showpacs,superscriptaddress,twocolumn]{revtex4-2}
\usepackage{amssymb,amsmath,xcolor,graphicx,psfrag}

\usepackage{hyperref, cancel}

\begin{document}

\newcommand{\jav}[1]{#1}

\title{Correlations at PT-symmetric quantum critical point}

\author{Bal\'azs D\'ora}
\email{dora.balazs@ttk.bme.hu}
\affiliation{MTA-BME Lend\"ulet Topology and Correlation Research Group,
Budapest University of Technology and Economics, 1521 Budapest, Hungary}
\affiliation{Department of Theoretical Physics, Budapest University of Technology and Economics, 1521 Budapest, Hungary}
\author{Doru Sticlet}
\affiliation{National Institute for R\&D of Isotopic and Molecular Technologies, 67-103 Donat, 400293 Cluj-Napoca, Romania}
\author{C\u{a}t\u{a}lin Pa\c{s}cu Moca}
\affiliation{MTA-BME Quantum Dynamics and Correlations Research Group, Institute of Physics, Budapest University of Technology and Economics, 1521 Budapest, Hungary}
\affiliation{Department  of  Physics,  University  of  Oradea,  410087,  Oradea,  Romania}

\date{\today}

\begin{abstract}
We consider a PT-symmetric Fermi gas with an exceptional point, representing the critical point between PT-symmetric and symmetry broken phases.
The low energy spectrum remains linear in momentum \jav{and is identical to that of a hermitian Fermi gas.}
The fermionic Green's function decays in a power law fashion for large distances, as expected from gapless excitations, albeit the exponent
is reduced from $-1$  due to the quantum Zeno effect.
In spite of the gapless nature of the excitations, the ground state entanglement entropy saturates to a finite value, independent of the subsystem size due to the
non-hermitian correlation length intrinsic to the system.
Attractive or repulsive interaction drives  the system into the PT-symmetry broken regime or opens up a gap and protects PT-symmetry, respectively.
Our results challenge the concept of universality in non-hermitian systems, where quantum criticality  can be masked due to non-hermiticity.
\end{abstract}

\maketitle

\paragraph{Introduction.}
Quantum criticality and universality play a prominent role in various branches of physics\cite{sachdev,herbutbook,continentino}, ranging from electrons in solids through
ultracold atoms to
 quantum plasma of quarks.
The emerging scale-invariance dictates 
the dispersion of excitation spectrum at the quantum critical point or the collapse of the gap upon approaching criticality, 
 which in turn determine the long distance behaviour of real space correlation functions. 
The influence of the critical point extends over a wide temperature window and its  universality applies not only in equilibrium, but 
extends also to near-adiabatic processes, such as the celebrated Kibble-Zurek mechanism\cite{kibble,zurek}.

Recently, non-hermitian systems have been extensively investigated\cite{gao2015,rotter,zeuner,Feng2014,hodaei,Bergholtz2021,ashidareview,ElGanainy2018,fruchart}, featuring, among many others, distinct kind of criticality and symmetry breakings.
PT-symmetric non-hermitian Hamiltonians\cite{mostafazadeh2002,mostafazadeh2003,Bender2007} possess either real or complex pairs of  eigenvalues, corresponding to eigenstates preserving or breaking PT-symmetry.
The transition from real to complex spectrum and the associated PT-symmetry breaking occurs at a non-hermitian quantum critical point, which is an exceptional point (EP)\cite{heiss}.
Therein, not only the spectrum becomes degenerate but also two (or more) \emph{eigenstates} coalesce, which then no longer form a complete basis.

In light of these, it would be important and interesting to explore to what extent the universality of
non-hermitian quantum critical points parallel their hermitian counterparts. In order to shed light on this issue, we  study a PT-symmetric Fermi gas, tuned to a PT-symmetric quantum critical point. Similar systems can be realized in various experiments.
We find that in spite of the gapless nature of low energy excitations, 
the spatial decay of the Green's function is faster at long distances than in hermitian systems, and cannot be accounted for by the critical exponents of the PT-symmetric EP.
The overlap of the hermitian and PT-symmetric ground states resembles the fidelity near a quantum critical point. 
The entanglement entropy saturates to a finite value with increasing subsystem size as if a finite gap was present in the system. All these features can be explained by measurement induced quantum Zeno effect\cite{misra,barontini},  arresting the propagation of correlations due to the underlying continuous measurement, required for non-hermitian dynamics.
Interaction effect are also subtle: repulsive or attractive interactions open up a gap and protect PT-symmetry or destroy PT-symmetry and induce a second order EP, respectively.

\paragraph{Hamiltonian.}
We study a one-dimensional \jav{ (direction $x$)}, PT-symmetric non-hermitian Hamiltonian, which reads as 
\begin{gather}
H=\int dx ~iv \left(R^+(x)\partial_xR(x)-L^+(x)\partial_xL(x)\right)+\nonumber\\
+\Delta R^+(x)L(x),
\label{hamilton}
\end{gather}
where $R(x)$ and $L(x)$ describe right and left moving fermionic fields\cite{giamarchi,nersesyan}, respectively, \jav{the system is half filled} and we assume $\Delta>0$ without loss of generality.
This Hamiltonian is apparently non-hermitian due to the absence of $\Delta L^+(x)R(x)$ term, 
and also resembles to a charge density wave\cite{nersesyan,gruner} Hamiltonian with only
half the couplings. \jav{Parity (P) transforms $[R(x,t),L(x,t)] \rightarrow [L(-x,t),R(-x,t)]$ while time reversal (T) results in 
$[R(x,t),L(x,t)]\rightarrow [L(x,-t),R(x,-t)]$ and 
takes the complex conjugate of complex numbers\cite{bender2005}. This Hamiltonian can be realized using an effective Lindblad description,
as discussed below Eq. \eqref{hamiltontb}.}

\jav{Eq. \eqref{hamilton} has  real eigenvalues}, which  follows from rewriting it in momentum space for a given momentum mode as
\begin{gather}
H_p=\left[\begin{array}{cc}
vp & \Delta\\
0 & -vp
\end{array}\right],
\label{hp}
\end{gather}
whose spectrum is purely real as $\varepsilon_\pm(p)=\pm v|p|$, independent from $\Delta$. It thus
describes  a critical system with dynamical critical exponent  $z=1$. Therefore, \jav{the partition function Tr$[\exp(-H/T)]$ is formally } 
identical to that in a hermitian Fermi gas.
In the following, we show that in spite of this, many physical 
properties are affected by the presence of $\Delta$ through a \emph{correlation length} associated to  $v/\Delta$.

Since $H_p$ is PT-symmetric and possesses a real spectrum, 
it can be brought to hermitian form after a similarity transformation\cite{mostafazadeh2002,mostafazadeh2003,bender2005}
as
\begin{gather}
\mathcal{S}^{-1}H_p\mathcal{S}=\left[\begin{array}{cc}
vp & 0\\
0 & -vp
\end{array}\right],\hspace*{5mm}
\mathcal{S}=\left[\begin{array}{cc}
1 & -\frac{\Delta}{2vp}\\
0 & 1
\end{array}\right].
\end{gather}

In Eq. \eqref{hp}, the $p=0$ point correspond to an exceptional point (EP)\cite{heiss,Bergholtz2021}, when the Hamiltonian becomes defective.
The $H_p$ is tuned to the brink of PT-symmetry breaking: by replacing the 0 with $\Gamma$, the spectrum changes to $\pm\sqrt{v^2p^2+\Gamma\Delta}$.
In Eq. \eqref{hamilton}, this amounts to adding $\Gamma \int dx~ L^+(x)R(x)$ to the Hamiltonian.
For $\Gamma>0$, a finite gap opens, preserving PT-symmetry while for $\Gamma<0$, a second order 
EP\cite{heiss} is induced as $\pm\sqrt{v^2p^2-|\Gamma|\Delta}$, and PT-symmetry is broken. 

Since the spectrum of Eq. \eqref{hamilton} is real, we can construct the ground state of the system as the minimal 
energy configuration\cite{EPAPS}, similarly to the hermitian realm~\footnote{Since the similarity
transformation is at our disposal, we can in principle calculate physical quantities either in the original non-hermitian setting or
by using the similarity transformation\cite{bender2006}.}.

\paragraph{Green's functions.}
We start analyzing the peculiar properties of Eq. \eqref{hamilton} by evaluating the equal time fermionic Green's functions, which are the 
normal $G(x)\equiv \langle R^+(x)R(0)\rangle$ and anomalous $F(x)\equiv\langle R^+(x)L(0)\rangle$ propagators. 
The conventional one reads\cite{EPAPS} as
\begin{gather}
G(x)=\int\limits_{0}^\infty\frac{dp}{2\pi}\frac{\Delta^2 e^{ipx-\alpha p}}{(2vp)^2+\Delta^2}+\int\limits_{-\infty}^0 \frac{dp}{2\pi}e^{ipx+\alpha p},
\label{gx}
\end{gather}
where the first term arises from the non-hermitian coupling between the right and left movers and is absent for $\Delta=0$, while the second term represent the conventional right-moving propagator\cite{delft} and
is responsible for the $1/x$ decay of the non-interacting hermitian Fermi gas. 
Here, we introduced the momentum space cutoff $\exp(-\alpha |p|)$ for momentum integrals\cite{giamarchi} with $\alpha$ the short distance cutoff.
\jav{In Eq. \eqref{gx}, the time dependence drops out completely since the spectrum is real.} 
In our case,  we find that
\begin{gather}
G(x)=-\frac{i}{2\pi x}\times\left\{\begin{array}{cc}
1 & x\ll v/\Delta\\
-2\left(\frac{2v}{\Delta x}\right)^2& v/\Delta\ll x
\end{array}\right. .
\end{gather}
Most importantly, we observe that in spite of the gapless nature of excitations\cite{continentino} with $z=1$ and the complete absence of $\Delta$ in the single particle spectrum, 
the real space Green's function decays as $x^{-3}$ beyond the non-hermitian correlation length as opposed to the $x^{-1}$ decay of free hermitian fermions.
This is interpreted as the quantum Zeno effect\cite{misra,ashidareview}, where the propagation of correlations is arrested by the continuous measurement within non-hermitian quantum mechanics, thus
significantly reduced correlations are present in the system in the end.
We note that within the correlation length, the conventional fermionic decay is revealed. Similar conclusions apply to $\langle L^+(x)L(0)\rangle$ correlator.

The decay of $G(x)$ is analogous to that in gapped or pseudogapped systems, where correlation functions decay differently within and outside of the correlation length.
However, in that case, an additional energy scale makes its presence felt already in the excitation spectrum, while this feature is completely missing in our non-hermitian
scenario, its spectrum $\pm v|p|$ is featureless.

Since the right and left moving fields are coupled by $\Delta$ in Eq. \eqref{hamilton}, an anomalous Green's function is also present, 
similarly to density waves\cite{gruner}, as
\begin{gather}
F(x)=-\int\limits_0^\infty \frac{dp}{2\pi} \frac{2\Delta vp e^{ipx-\alpha p}}{(2vp)^2+\Delta^2},
\end{gather}
exhibiting
\begin{gather}
F(x)=\frac{\Delta }{4\pi v}\times\left\{\begin{array}{cc}
\ln\frac{\Delta |x|e^\gamma}{2v} 
& x\ll v/\Delta\\
\left(\frac{2v}{\Delta x}\right)^2 & v/\Delta\ll x
\end{array}\right. 
\end{gather}
behaviour, $\gamma\approx 0.5772$ is Euler's constant\cite{gradstein}, and $F(0)=\frac{\Delta}{4\pi v}\ln\frac{\Delta\alpha e^\gamma}{2v}$.
In a conventional gapped density wave, this correlation function decays exponentially as $\exp(-|x|\Delta/v)$. In contrast, a long distance 
power law decay is identified here
due to gapless excitations. 
\jav{Since $F(x)$ is directly proportional to $\Delta$ and vanishes in a hermitian Fermi gas, it can be used to reveal the presence of $\Delta$ 
at short distances too.}
The correlator $\langle L^+(x)R(0)\rangle$ follows similar behaviour in spite of the fact that no $L^+R$ coupling 
is present in the Hamiltonian.
By moving away from the critical point, the Green's function decays exponentially deep in the PT-symmetric phase with an exponent reduced by the quantum Zeno effect,
while in the PT-broken regime, a power law decay $\sim x^{-3/2}$ shows up due to the second order EP.

The ground state carries a finite current 
 since the non-hermitian term in Eq. \eqref{hamilton} converts left to right movers. The particle current density\cite{nersesyan,giamarchi} $j$ is
\begin{gather}
j=v \left\langle R^+(x)R(x)-L^+(x)L(x)\right\rangle =\frac{\Delta}{4},
\end{gather}
independent of $x$.
At the same time, our system realizes a charge density wave as the real space density profile, $n(x)$ oscillates as
\begin{gather}
n(x)-n_0=\frac{\Delta}{2\pi v}\ln\left(\frac{\Delta\alpha e^\gamma}{2v}\right)\cos(2k_Fx),
\end{gather}
where $n_0$ is the homogeneous particle density in the $\Delta=0$ system and $k_F$ is the Fermi wavenumber.

\paragraph{Fidelity.}
To investigate the relation between the gapless hermitian ($\Delta=0$) and PT-symmetric ($\Delta\neq 0$) ground states, their overlap, the fidelity\cite{nielsen,rams},
 is evaluated to yield  
\begin{gather}
\langle\Psi_0|\Psi_\Delta\rangle=\frac{1}{\sqrt{\cosh(L\Delta/4v)}}=\left\{\begin{array}{cc}
e^{-\frac{L\Delta}{8v}+\frac{\ln 2}{2}} & \frac{L\Delta}{v}\gg 1 \\
1-\left(\frac{L\Delta}{8 v}\right)^2 & \frac{L\Delta}{v}\ll 1
\end{array}\right.
\end{gather}
This is analogous to the behaviour expected in hermitian systems around a hermitian quantum critical point\cite{rams}, although in the present case, both
systems with $\Delta=0$ and $\Delta\neq 0$ are identically critical with identical gapless spectra. 
This supports the narrative associated to non-hermitian correlation length, masking
the gaplessness of the spectrum in various physical quantities.

\paragraph{Lattice realization.}
Our system can be realized with fermions in a one-dimensional half-filled tight binding chain \jav{with periodic boundary condition (PBC) and even number of sites} as
\begin{gather}
H^{tb}=\sum_{n}\frac{J+i\delta(-1)^n}{2}\left(c^+_nc_{n+1}+c^+_{n+1}c_n\right)+(-1)^n g c^+_nc_n,
\label{hamiltontb}
\end{gather}
where $J$ is the uniform hopping, $g$ is an alternating on-site potential, $\delta$ stems from a non-hermitian alternating hopping and $c$'s are fermionic 
annihilation operators. \jav{The non-hermitian term arises from an effective Lindblad-equation without the recycling term\cite{daley}
using jump operators for bonds\cite{ashida2018,gongprx,ashidasinegordon,takasu} 
as $\sqrt{\delta}(c_n\pm c_{n+1})$ on even (+) and odd (-) bonds, and neglecting a constant shift, proportional to the total particle number.}
In momentum space, the ensuing Hamiltonian in the basis of $(c_k,c_{k-\pi})$ with $0\leqslant k<\pi$ reads as
\begin{gather}
H^{tb}_k=\left[\begin{array}{cc}
J\cos(k) & g+\delta\sin(k)\\
g-\delta\sin(k) & -J\cos(k)
\end{array}\right]
\end{gather}
\jav{with dispersion $E_{\pm}(k)=\pm\sqrt{(J^2+\delta^2)\cos^2(k)+g^2-\delta^2}$.}
Expanding this around the Fermi wavenumber $\pi/2$ for half filling, we get Eq. \eqref{hamilton} for $g=\delta$ as $v=J$ and $\Delta=2g$.
As long as $g\geq \delta$, the system is PT-symmetric and the spectrum is real. However, for $g<\delta$, there is always a region close 
to $k=\pi/2$ with imaginary pairs of eigenvalues,
thus PT-symmetry is broken. From this on, we consider the $g=\delta$ case, when $H^{tb}$ is tuned to the boundary of PT symmetry and 
represents a lattice realization of Eq. \eqref{hamilton}. For more details of $H^{tb}$, see Ref. \cite{EPAPS}.

\begin{figure}[h!]
\centering
\includegraphics[width=7cm]{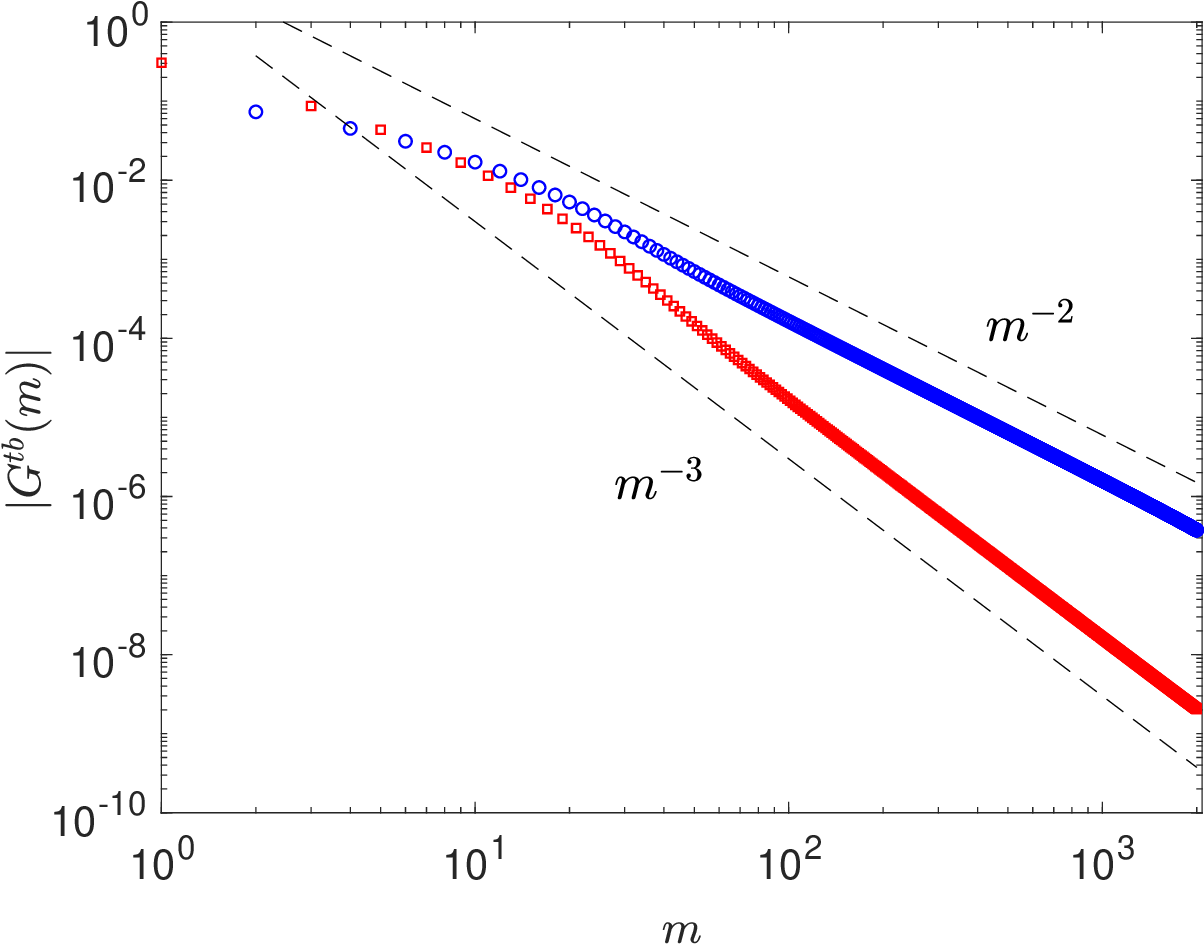}
\caption{The absolute value of the lattice Green's function is plotted for even (blue) and odd (red) spatial separation and $g=\delta=0.2J$. The reduction caused by the 
quantum Zeno effect is clearly visible in the long distance asymptotics, which agrees with the continuum limit calculation. The black dashed lines show the $m^{-2}$ and 
$m^{-3}$ power law decays.}
\label{greentb}
\end{figure}

First,
we evaluate numerically the fermionic Green's function, $G^{tb}(m)=\langle c^+_{m+n}c_n\rangle $. Due to the commensurate Fermi wavenumber $\pi/2$, 
this measures directly $G(x)$ or $F(x)$ for 
odd or even $n$, because they are multiplied by the $\sin(k_Fm)$ and $\cos(k_Fm)$ factors in the full fermionic Green's function, respectively. 
In particular, for $g=\delta=0$, we recover the conventional expression $G^{tb}(m)=\sin(\pi n/2)/n\pi$.
For finite $g=\delta$, the system  exhibits indeed charge density wave pattern with gapless, linearly dispersing excitations.
In Fig. \ref{greentb}, we numerically evaluate the tight binding Green's function in the thermodynamic limit, which agree with $G(x)$ and $F(x)$, as advertised above.

\begin{figure}[t!]
\centering
\includegraphics[width=7cm]{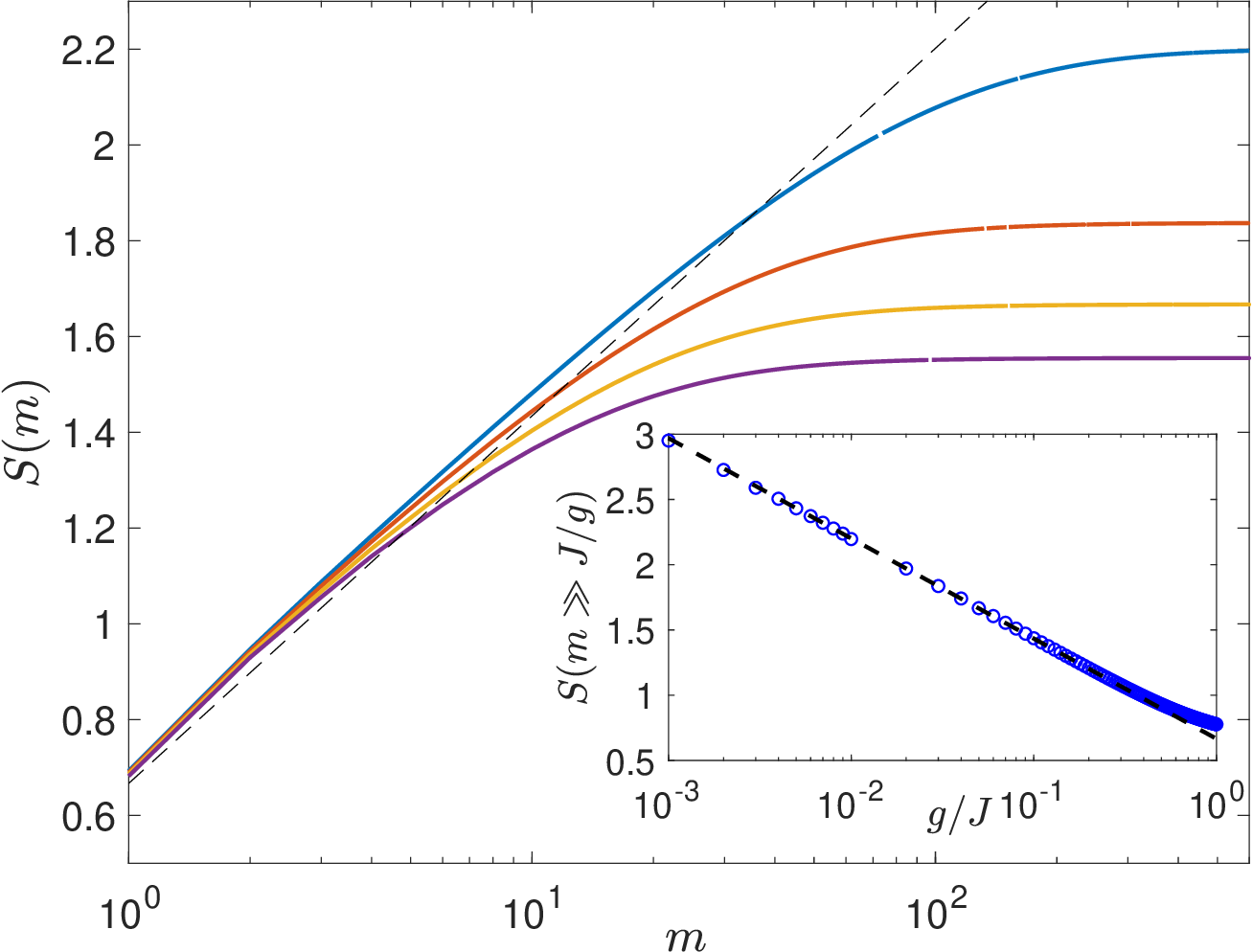}
\caption{The von-Neumann entanglement entropy is plotted for $g=\delta=0.01J$, $0.03J$, $0.05J$ and $0.07J$ from top to bottom. The black dashed line denotes $\frac 13\ln(m)+\frac 23$ curve. The inset shows 
the saturation value of the entanglement entropy for large subsystem size as a function of $g=\delta$. The black dashed line denotes $\frac 13\ln(J/g)+\frac 23$.}
\label{entropyep}
\end{figure}

\paragraph{Entanglement.}
With the knowledge of the single particle Green's function, we  address the entanglement properties of our system.
In particular, we evaluate the von-Neumann entanglement entropy, $S(m)$ between the subsystem of size $m$ and the rest of the chain using Refs. \onlinecite{Peschel2009,herviou,chang20,maity}. We have also checked
on small systems up to 26 sites that by numerically exact diagonalizing the many-body problem of Eq. \eqref{hamiltontb} and brute force calculating the subsystem entanglement agrees with
the Green's function based approach.
We find that for small subsystem size $m$, the entanglement entropy follows a $S(m)=\frac{1}{3}\ln(m)$ growth\cite{calabrese2004,Pollmann-2009}. This indicates that the central charge\cite{eisert,amico, nielsen,srednicki} 
of the non-hermitian system
with EP in Eq. \eqref{hamilton} remains 1. Upon further increasing the subsystem size such that $m\gg J/g$, the entanglement entropy saturates to a constant value, shown in Fig. \ref{entropyep}, similarly to
what happens in one-dimensional gapped systems. However, our Hamiltonian is gapless, but the presence of non-hermitian coherence length stops the logarithmic
entanglement growth, similarly to how it affects the decay of the Green's function through the quantum Zeno effect\cite{misra}.
The scaling of the saturation value of the entanglement entropy for large subsystem  agrees surprisingly with Ref. \cite{calabrese2004} as $S(m\gg J/g)=\frac 13\ln(J/g)+\frac 23$ 
after identifying
the correlation length with the non-hermitian coupling. 
This behaviour is highly unusual and challenges the interpretation of criticality and universality in non-hermitian systems: in spite of the typical gapless excitation spectrum
in one dimension, the entanglement entropy still saturates as if the system was gapped. This "contradiction" is cured upon realizing that in spite of the gapless spectrum,
the asymptotic decay of the Green's function gets suppressed by the quantum Zeno effect, which results in an effective gapped type subsystem entanglement entropy.

\jav{We mention that in related  non-unitary conformal field theories and critical systems\cite{Bianchini14,Bianchini16,Couvreur,Dupic}, the conventional $S(m)\sim \ln(m)$ behaviour was found. The difference between these and our findings follows from 
the fact that either a distinct definition of the reduced density matrix is used\cite{Couvreur} or in the models of Refs. \cite{Bianchini14,Bianchini16,Dupic}
the right ground states of both $H$ and $H^+$ are equal to each other, which is not the case\cite{EPAPS} for Eq. \eqref{hp}.}

\paragraph{Interaction effects.}
Since low dimensional systems are susceptible to various  instabilities, we address here the role of electron-electron interactions on Eqs. \eqref{hamilton} and 
\eqref{hamiltontb}. On the lattice, this amounts to considering the nearest-neighbour interaction as
$H_{int}=V\sum_m c^+_{m}c_m c^+_{m+1}c_{m+1}$. This interaction preserves PT-symmetry.
In similar situations\cite{giamarchi}, one typically performs a renormalization group calculation to address the  relevance of interaction processes. In our case, however, 
already a perturbative Hartree-Fock calculation suffices as
\begin{gather}
H^{HF}_{int}=V\sum_m 2\langle c^+_{m+1}c_{m+1}\rangle c^+_{m}c_m-\nonumber\\
-\langle c^+_{m+1}c_m\rangle c^+_m c_{m+1}-\langle c^+_{m}c_{m+1}\rangle c^+_{m+1}c_{m}+\textmd{const},
\label{inthf}
\end{gather}
where the expectation values are taken with respect to the non-interacting ground state. The first Hartree term on the r.h.s. renormalizes $g$ by $2Vg/(\pi J)\ln(2J/g)$,
while the last two Fock terms mostly induce a shift to the Fermi wavevector, which
 turns out to be irrelevant at weak interactions,
and the first term in Eq. \eqref{inthf} governs the weak coupling physics.
For repulsive interactions,
the renormalized $g$ is enhanced,
thus a clean gap opens up in the spectrum and protects PT-symmetry.
On the attractive side, $g$ gets renormalized to smaller values, hence $\delta$ becomes larger then the renormalized $g$, the PT-symmetry gets
broken and part of the excitation spectrum becomes imaginary.

\begin{figure}[h!]
\centering
\includegraphics[width=7cm]{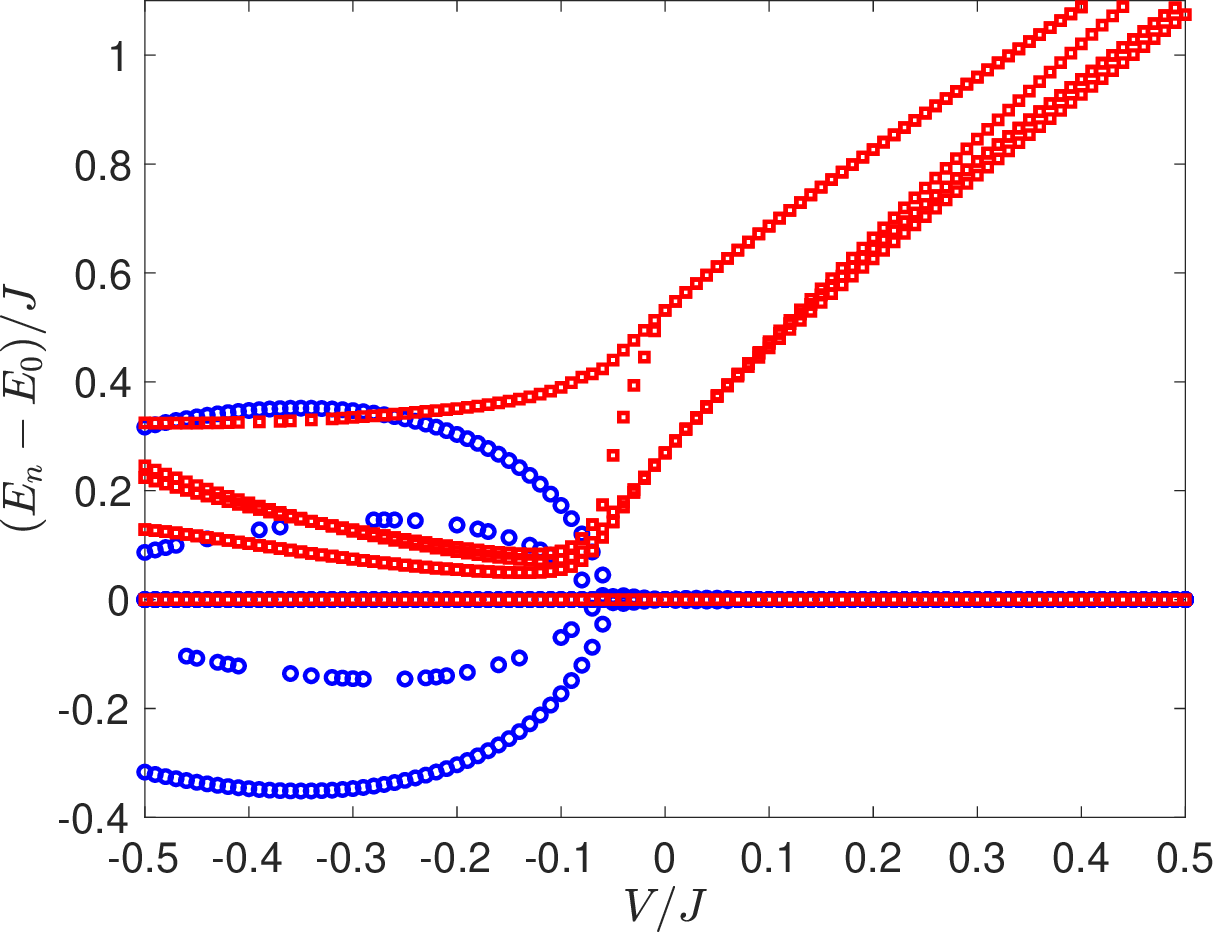}
\caption{The many-body spectrum $E_n$ of $H^{tb}+H_{int}$, consisting of the 6 lowest lying states above the 
\emph{real} ground state energy $E_0$ is shown for 26 lattice sites with $g=\delta=0.5J$. The real and imaginary part of the 
eigenenergies, measured from the ground state energy are plotted in red squares and blue circles, respectively. }
\label{manybodyspec}
\end{figure}

This conclusion is corroborated by studying $H^{tb}+H_{int}$ numerically using exact diagonalization on a system with 26 lattice sites and 13 particles \jav{and PBC}. 
We study the spectrum around the ground state energy $E_0$ for weak interactions, where $E_0$ represents the eigenenergy with the lowest real part. 
This turns out to be unique and purely \emph{real} in this case. The many-body
spectrum for the 6 lowest lying states above the ground state are plotted in Fig. \ref{manybodyspec}.
There is a finite gap 
in the non-interacting, $V=0$ limit due to finite level spacing, therefore the PT-symmetry breaking is shifted to finite negative $V$. This together with the 
 finite non-interacting gap vanishes with increasing system size.

\paragraph{Relation to sine-Gordon model.}
One-dimensional quantum systems are often studied via bosonization\cite{giamarchi}. By applying it to Eq. \eqref{hamilton}, we get
\begin{gather}
H=\int \frac{dx}{2\pi} v\left[(\pi\Pi(x))^2+(\partial_x\phi(x))^2\right]+\frac{\Delta}{2\pi\alpha}e^{-i2\phi(x)},
\label{hamsg}
\end{gather}
where $\Pi$ and $\phi$ are dual fields satisfying $[\phi(x_1),\Pi(x_2)]=i\delta(x_1-x_2)$ \jav{at equal times}. This is the non-hermitian "half" of the 
conventional sine-Gordon model and 
both the imaginary sine and real cosine potentials
have equal strengths\cite{bender2005,ashidasinegordon}. Due to this, the absolute value of the potential term is constant and is unable to localize the $\phi$ field.
This is manifested in the linear energy-momentum relationship from the original, fermionic treatment. The fate of adding $e^{i2\phi(x)}$ potential to 
Eq. \eqref{hamsg} is discussed in Ref. \cite{EPAPS}.
We  speculate that a mapping of this bosonic Hamiltonian back into the
fermionic one could play an important role when analyzing strongly correlated non-hermitian systems\cite{takasu} at their respective Luther-Emery point\cite{lutheremery}.

\paragraph{Experimental possibilities.}
Eqs. \eqref{hamilton} and \eqref{hamiltontb} can be realized using a waveguide lattice\cite{Song2020}, simulating an effective non-hermitian Hamiltonian experimentally. 
From Refs. \cite{Lamata2007,Gerritsma2010,Lee2015}, our setup can be created with ion trap physics, simulating  our system in a minimal setting consisting of two atomic levels 
and a motional degree of freedom. Ultracold bosonic atoms loaded into an optical lattice as in Refs. \cite{takasu,ashidasinegordon} with controlled losses
represent another viable route.
In this case, the density matrix of the system evolves according to a 
Lindblad equation. By continuously measuring the environment\cite{daley,carmichael,ashidareview} and postselecting the data to ensure that the state of the environment remains unchanged, 
the effective Hamiltonian becomes non-hermitian  and
of the form Eq. \eqref{hamilton}. 
Additionally, one can use single photon interferometry to realize our momentum space Hamiltonians\cite{PRXQ}.

\paragraph{Summary.}
We have studied the properties of a PT-symmetric quantum critical point. We find that it is "less universal" than its hermitian counterpart: the equal time Green's function decays faster for long distances than expected from the linear energy-momentum relationship. This occurs due to the quantum Zeno effect, which
slows down the propagation of excitations due to continuous measurement. As a result, the spatial entanglement entropy saturates 
to a finite value with increasing subsystem size, and the saturation value is determined by the non-hermitian correlation length, in spite of the gapless nature of excitations.
Repulsive electron-electron interaction opens up a gap by protecting PT-symmetry while attractive interaction breaks PT-symmetry and induces a second order exceptional point.

\begin{acknowledgments}
A useful exchange of e-mails with Y. Ashida is gratefully acknowledged.
This research is supported by the National Research, Development and 
Innovation Office - NKFIH  within the Quantum Technology National Excellence 
Program (Project No.~2017-1.2.1-NKP-2017-00001), K134437, by the BME-Nanotechnology 
FIKP grant (BME FIKP-NAT), and by a grant of the Ministry of Research, Innovation and
 Digitization, CNCS/CCCDI-UEFISCDI, under projects number PN-III-P4-ID-PCE-2020-0277 and  PN-III-P1-1.1-TE-2019-0423, within PNCDI III.
\end{acknowledgments}

\bibliographystyle{apsrev}
\bibliography{wboson1}

\begin{thebibliography}{10}
\expandafter\ifx\csname bibnamefont\endcsname\relax
  \def\bibnamefont#1{#1}\fi
\expandafter\ifx\csname bibfnamefont\endcsname\relax
  \def\bibfnamefont#1{#1}\fi
\expandafter\ifx\csname url\endcsname\relax
  \def\url#1{\texttt{#1}}\fi
\expandafter\ifx\csname urlprefix\endcsname\relax\def\urlprefix{URL }\fi
\providecommand{\bibinfo}[2]{#2}
\providecommand{\eprint}[2][]{\url{#2}}

\bibitem{sachdev}
\bibinfo{author}{\bibfnamefont{S.}~\bibnamefont{Sachdev}},
  \emph{\bibinfo{title}{Quantum Phase Transitions}}
  (\bibinfo{publisher}{Cambridge Univ. Press}, \bibinfo{address}{Cambridge},
  \bibinfo{year}{1999}).

\bibitem{herbutbook}
\bibinfo{author}{\bibfnamefont{I.}~\bibnamefont{Herbut}},
  \emph{\bibinfo{title}{A Modern Approach to Critical Phenomena}}
  (\bibinfo{publisher}{Cambridge University Press}, \bibinfo{year}{2007}).

\bibitem{continentino}
\bibinfo{author}{\bibfnamefont{M.}~\bibnamefont{Continentino}},
  \emph{\bibinfo{title}{Quantum Scaling in Many-Body Systems: An Approach to
  Quantum Phase Transitions}} (\bibinfo{publisher}{Cambridge University Press},
  \bibinfo{year}{2017}), \bibinfo{edition}{2nd} ed.

\bibitem{kibble}
\bibinfo{author}{\bibfnamefont{T.~W.~B.} \bibnamefont{Kibble}},
  \emph{\bibinfo{title}{Topology of cosmic domains and strings}},
  \bibinfo{journal}{J. Phys. A} \textbf{\bibinfo{volume}{9}},
  \bibinfo{pages}{1387} (\bibinfo{year}{1976}).

\bibitem{zurek}
\bibinfo{author}{\bibfnamefont{W.~H.} \bibnamefont{Zurek}},
  \emph{\bibinfo{title}{Cosmological experiments in superfluid helium?}},
  \bibinfo{journal}{Nature} \textbf{\bibinfo{volume}{317}},
  \bibinfo{pages}{505} (\bibinfo{year}{1985}).

\bibitem{gao2015}
\bibinfo{author}{\bibfnamefont{T.}~\bibnamefont{Gao}},
  \bibinfo{author}{\bibfnamefont{E.}~\bibnamefont{Estrecho}},
  \bibinfo{author}{\bibfnamefont{K.~Y.} \bibnamefont{Bliokh}},
  \bibinfo{author}{\bibfnamefont{T.~C.~H.} \bibnamefont{Liew}},
  \bibinfo{author}{\bibfnamefont{M.~D.} \bibnamefont{Fraser}},
  \bibinfo{author}{\bibfnamefont{S.}~\bibnamefont{Brodbeck}},
  \bibinfo{author}{\bibfnamefont{M.}~\bibnamefont{Kamp}},
  \bibinfo{author}{\bibfnamefont{C.}~\bibnamefont{Schneider}},
  \bibinfo{author}{\bibfnamefont{S.}~\bibnamefont{H{\"o}fling}},
  \bibinfo{author}{\bibfnamefont{Y.}~\bibnamefont{Yamamoto}},
  \bibinfo{author}{\bibfnamefont{F.}~\bibnamefont{Nori}},
  \bibinfo{author}{\bibfnamefont{Y.~S.} \bibnamefont{Kivshar}}, \emph{et~al.},
  \emph{\bibinfo{title}{Observation of non-hermitian degeneracies in a chaotic
  exciton-polariton billiard}}, \bibinfo{journal}{Nature}
  \textbf{\bibinfo{volume}{526}}, \bibinfo{pages}{554} (\bibinfo{year}{2015}).

\bibitem{rotter}
\bibinfo{author}{\bibfnamefont{I.}~\bibnamefont{Rotter}} \bibnamefont{and}
  \bibinfo{author}{\bibfnamefont{J.~P.} \bibnamefont{Bird}},
  \emph{\bibinfo{title}{A review of progress in the physics of open quantum
  systems: theory and experiment}}, \bibinfo{journal}{Rep. Prog. Phys.}
  \textbf{\bibinfo{volume}{78}}, \bibinfo{pages}{114001}
  (\bibinfo{year}{2015}).

\bibitem{zeuner}
\bibinfo{author}{\bibfnamefont{J.~M.} \bibnamefont{Zeuner}},
  \bibinfo{author}{\bibfnamefont{M.~C.} \bibnamefont{Rechtsman}},
  \bibinfo{author}{\bibfnamefont{Y.}~\bibnamefont{Plotnik}},
  \bibinfo{author}{\bibfnamefont{Y.}~\bibnamefont{Lumer}},
  \bibinfo{author}{\bibfnamefont{S.}~\bibnamefont{Nolte}},
  \bibinfo{author}{\bibfnamefont{M.~S.} \bibnamefont{Rudner}},
  \bibinfo{author}{\bibfnamefont{M.}~\bibnamefont{Segev}}, \bibnamefont{and}
  \bibinfo{author}{\bibfnamefont{A.}~\bibnamefont{Szameit}},
  \emph{\bibinfo{title}{Observation of a topological transition in the bulk of
  a non-hermitian system}}, \bibinfo{journal}{Phys. Rev. Lett.}
  \textbf{\bibinfo{volume}{115}}, \bibinfo{pages}{040402}
  (\bibinfo{year}{2015}).

\bibitem{Feng2014}
\bibinfo{author}{\bibfnamefont{L.}~\bibnamefont{Feng}},
  \bibinfo{author}{\bibfnamefont{Z.~J.} \bibnamefont{Wong}},
  \bibinfo{author}{\bibfnamefont{R.-M.} \bibnamefont{Ma}},
  \bibinfo{author}{\bibfnamefont{Y.}~\bibnamefont{Wang}}, \bibnamefont{and}
  \bibinfo{author}{\bibfnamefont{X.}~\bibnamefont{Zhang}},
  \emph{\bibinfo{title}{Single-mode laser by parity-time symmetry breaking}},
  \bibinfo{journal}{Science}
  \textbf{\bibinfo{volume}{346}}(\bibinfo{number}{6212}), \bibinfo{pages}{972}
  (\bibinfo{year}{2014}).

\bibitem{hodaei}
\bibinfo{author}{\bibfnamefont{H.}~\bibnamefont{Hodaei}},
  \bibinfo{author}{\bibfnamefont{A.~U.} \bibnamefont{Hassan}},
  \bibinfo{author}{\bibfnamefont{S.}~\bibnamefont{Wittek}},
  \bibinfo{author}{\bibfnamefont{H.}~\bibnamefont{Garcia-Gracia}},
  \bibinfo{author}{\bibfnamefont{R.}~\bibnamefont{El-Ganainy}},
  \bibinfo{author}{\bibfnamefont{D.~N.} \bibnamefont{Christodoulides}},
  \bibnamefont{and}
  \bibinfo{author}{\bibfnamefont{M.}~\bibnamefont{Khajavikhan}},
  \emph{\bibinfo{title}{Enhanced sensitivity at higher-order exceptional
  points}}, \bibinfo{journal}{Nature} \textbf{\bibinfo{volume}{548}},
  \bibinfo{pages}{187} (\bibinfo{year}{2017}).

\bibitem{Bergholtz2021}
\bibinfo{author}{\bibfnamefont{E.~J.} \bibnamefont{Bergholtz}},
  \bibinfo{author}{\bibfnamefont{J.~C.} \bibnamefont{Budich}},
  \bibnamefont{and} \bibinfo{author}{\bibfnamefont{F.~K.} \bibnamefont{Kunst}},
  \emph{\bibinfo{title}{Exceptional topology of non-hermitian systems}},
  \bibinfo{journal}{Rev. Mod. Phys.} \textbf{\bibinfo{volume}{93}},
  \bibinfo{pages}{015005} (\bibinfo{year}{2021}).

\bibitem{ashidareview}
\bibinfo{author}{\bibfnamefont{Y.}~\bibnamefont{Ashida}},
  \bibinfo{author}{\bibfnamefont{Z.}~\bibnamefont{Gong}}, \bibnamefont{and}
  \bibinfo{author}{\bibfnamefont{M.}~\bibnamefont{Ueda}},
  \emph{\bibinfo{title}{Non-hermitian physics}}, \bibinfo{journal}{Advances in
  Physics} \textbf{\bibinfo{volume}{69}}, \bibinfo{pages}{3}
  (\bibinfo{year}{2020}).

\bibitem{ElGanainy2018}
\bibinfo{author}{\bibfnamefont{R.}~\bibnamefont{El-Ganainy}},
  \bibinfo{author}{\bibfnamefont{K.~G.} \bibnamefont{Makris}},
  \bibinfo{author}{\bibfnamefont{M.}~\bibnamefont{Khajavikhan}},
  \bibinfo{author}{\bibfnamefont{Z.~H.} \bibnamefont{Musslimani}},
  \bibinfo{author}{\bibfnamefont{S.}~\bibnamefont{Rotter}}, \bibnamefont{and}
  \bibinfo{author}{\bibfnamefont{D.~N.} \bibnamefont{Christodoulides}},
  \emph{\bibinfo{title}{Non-hermitian physics and pt symmetry}},
  \bibinfo{journal}{Nat. Phys.}
  \textbf{\bibinfo{volume}{14}}(\bibinfo{number}{1}), \bibinfo{pages}{11}
  (\bibinfo{year}{2018}).

\bibitem{fruchart}
\bibinfo{author}{\bibfnamefont{M.}~\bibnamefont{Fruchart}},
  \bibinfo{author}{\bibfnamefont{R.}~\bibnamefont{Hanai}},
  \bibinfo{author}{\bibfnamefont{P.~B.} \bibnamefont{Littlewood}},
  \bibnamefont{and} \bibinfo{author}{\bibfnamefont{V.}~\bibnamefont{Vitelli}},
  \emph{\bibinfo{title}{Non-reciprocal phase transitions}},
  \bibinfo{journal}{Nature} \textbf{\bibinfo{volume}{592}},
  \bibinfo{pages}{363} (\bibinfo{year}{2021}).

\bibitem{mostafazadeh2002}
\bibinfo{author}{\bibfnamefont{A.}~\bibnamefont{Mostafazadeh}},
  \emph{\bibinfo{title}{Pseudo-hermiticity versus pt symmetry: The necessary
  condition for the reality of the spectrum of a non-hermitian hamiltonian}},
  \bibinfo{journal}{Journal of Mathematical Physics}
  \textbf{\bibinfo{volume}{43}}(\bibinfo{number}{1}), \bibinfo{pages}{205}
  (\bibinfo{year}{2002}).

\bibitem{mostafazadeh2003}
\bibinfo{author}{\bibfnamefont{A.}~\bibnamefont{Mostafazadeh}},
  \emph{\bibinfo{title}{{Exact PT}-symmetry is equivalent to hermiticity}},
  \bibinfo{journal}{Journal of Physics A: Mathematical and General}
  \textbf{\bibinfo{volume}{36}}(\bibinfo{number}{25}), \bibinfo{pages}{7081}
  (\bibinfo{year}{2003}).

\bibitem{Bender2007}
\bibinfo{author}{\bibfnamefont{C.~M.} \bibnamefont{Bender}},
  \emph{\bibinfo{title}{Making sense of non-hermitian hamiltonians}},
  \bibinfo{journal}{Reports on Progress in Physics}
  \textbf{\bibinfo{volume}{70}}(\bibinfo{number}{6}), \bibinfo{pages}{947}
  (\bibinfo{year}{2007}).

\bibitem{heiss}
\bibinfo{author}{\bibfnamefont{W.~D.} \bibnamefont{Heiss}},
  \emph{\bibinfo{title}{The physics of exceptional points}},
  \bibinfo{journal}{J. Phys. A: Math. Theor.}
  \textbf{\bibinfo{volume}{45}}(\bibinfo{number}{44}), \bibinfo{pages}{444016}
  (\bibinfo{year}{2012}).

\bibitem{misra}
\bibinfo{author}{\bibfnamefont{B.}~\bibnamefont{Misra}} \bibnamefont{and}
  \bibinfo{author}{\bibfnamefont{E.~C.~G.} \bibnamefont{Sudarshan}},
  \emph{\bibinfo{title}{The zeno’s paradox in quantum theory}},
  \bibinfo{journal}{Journal of Mathematical Physics}
  \textbf{\bibinfo{volume}{18}}(\bibinfo{number}{4}), \bibinfo{pages}{756}
  (\bibinfo{year}{1977}).

\bibitem{barontini}
\bibinfo{author}{\bibfnamefont{G.}~\bibnamefont{Barontini}},
  \bibinfo{author}{\bibfnamefont{R.}~\bibnamefont{Labouvie}},
  \bibinfo{author}{\bibfnamefont{F.}~\bibnamefont{Stubenrauch}},
  \bibinfo{author}{\bibfnamefont{A.}~\bibnamefont{Vogler}},
  \bibinfo{author}{\bibfnamefont{V.}~\bibnamefont{Guarrera}}, \bibnamefont{and}
  \bibinfo{author}{\bibfnamefont{H.}~\bibnamefont{Ott}},
  \emph{\bibinfo{title}{Controlling the dynamics of an open many-body quantum
  system with localized dissipation}}, \bibinfo{journal}{Phys. Rev. Lett.}
  \textbf{\bibinfo{volume}{110}}, \bibinfo{pages}{035302}
  (\bibinfo{year}{2013}).

\bibitem{giamarchi}
\bibinfo{author}{\bibfnamefont{T.}~\bibnamefont{Giamarchi}},
  \emph{\bibinfo{title}{Quantum Physics in One Dimension}}
  (\bibinfo{publisher}{Oxford University Press}, \bibinfo{address}{Oxford},
  \bibinfo{year}{2004}).

\bibitem{nersesyan}
\bibinfo{author}{\bibfnamefont{A.~O.} \bibnamefont{Gogolin}},
  \bibinfo{author}{\bibfnamefont{A.~A.} \bibnamefont{Nersesyan}},
  \bibnamefont{and} \bibinfo{author}{\bibfnamefont{A.~M.}
  \bibnamefont{Tsvelik}}, \emph{\bibinfo{title}{Bosonization and Strongly
  Correlated Systems}} (\bibinfo{publisher}{Cambridge University Press},
  \bibinfo{address}{Cambridge}, \bibinfo{year}{1998}).

\bibitem{gruner}
\bibinfo{author}{\bibfnamefont{G.}~\bibnamefont{Gr\"uner}},
  \emph{\bibinfo{title}{Density waves in solids}}
  (\bibinfo{publisher}{Addison-Wesley}, \bibinfo{address}{Reading},
  \bibinfo{year}{1994}).

\bibitem{bender2005}
\bibinfo{author}{\bibfnamefont{C.~M.} \bibnamefont{Bender}},
  \bibinfo{author}{\bibfnamefont{H.}~\bibnamefont{Jones}}, \bibnamefont{and}
  \bibinfo{author}{\bibfnamefont{R.}~\bibnamefont{Rivers}},
  \emph{\bibinfo{title}{Dual pt-symmetric quantum field theories}},
  \bibinfo{journal}{Physics Letters B}
  \textbf{\bibinfo{volume}{625}}(\bibinfo{number}{3}), \bibinfo{pages}{333}
  (\bibinfo{year}{2005}).

\bibitem{EPAPS}
\bibinfo{note}{See EPAPS Document No. XXX for supplementary material providing
  further technical details.}

\bibitem{Note1}
\bibinfo{note}{Since the similarity transformation is at our disposal, we can
  in principle calculate physical quantities either in the original
  non-hermitian setting or by using the similarity transformation\cite
  {bender2006}.}

\bibitem{delft}
\bibinfo{author}{\bibfnamefont{J.}~\bibnamefont{von Delft}} \bibnamefont{and}
  \bibinfo{author}{\bibfnamefont{H.}~\bibnamefont{Schoeller}},
  \emph{\bibinfo{title}{Bosonization for beginners -- refermionization for
  experts}}, \bibinfo{journal}{Ann. Phys. (Leipzig)}
  \textbf{\bibinfo{volume}{7}}, \bibinfo{pages}{225} (\bibinfo{year}{1998}).

\bibitem{gradstein}
\bibinfo{author}{\bibfnamefont{I.}~\bibnamefont{Gradshteyn}} \bibnamefont{and}
  \bibinfo{author}{\bibfnamefont{I.}~\bibnamefont{Ryzhik}},
  \emph{\bibinfo{title}{Table of Integrals, Series, and Products}}
  (\bibinfo{publisher}{Academic Press}, \bibinfo{address}{New York},
  \bibinfo{year}{2007}).

\bibitem{nielsen}
\bibinfo{author}{\bibfnamefont{M.}~\bibnamefont{Nielsen}} \bibnamefont{and}
  \bibinfo{author}{\bibfnamefont{I.}~\bibnamefont{Chuang}},
  \emph{\bibinfo{title}{Quantum Computation and Quantum Information}}
  (\bibinfo{publisher}{Cambridge University Press},
  \bibinfo{address}{Cambridge}, \bibinfo{year}{2000}).

\bibitem{rams}
\bibinfo{author}{\bibfnamefont{M.~M.} \bibnamefont{Rams}} \bibnamefont{and}
  \bibinfo{author}{\bibfnamefont{B.}~\bibnamefont{Damski}},
  \emph{\bibinfo{title}{Quantum fidelity in the thermodynamic limit}},
  \bibinfo{journal}{Phys. Rev. Lett.} \textbf{\bibinfo{volume}{106}},
  \bibinfo{pages}{055701} (\bibinfo{year}{2011}).

\bibitem{daley}
\bibinfo{author}{\bibfnamefont{A.~J.} \bibnamefont{Daley}},
  \emph{\bibinfo{title}{Quantum trajectories and open many-body quantum
  systems}}, \bibinfo{journal}{Advances in Physics}
  \textbf{\bibinfo{volume}{63}}, \bibinfo{pages}{77} (\bibinfo{year}{2014}).

\bibitem{ashida2018}
\bibinfo{author}{\bibfnamefont{Y.}~\bibnamefont{Ashida}} \bibnamefont{and}
  \bibinfo{author}{\bibfnamefont{M.}~\bibnamefont{Ueda}},
  \emph{\bibinfo{title}{Full-counting many-particle dynamics: Nonlocal and
  chiral propagation of correlations}}, \bibinfo{journal}{Phys. Rev. Lett.}
  \textbf{\bibinfo{volume}{120}}, \bibinfo{pages}{185301}
  (\bibinfo{year}{2018}).

\bibitem{gongprx}
\bibinfo{author}{\bibfnamefont{Z.}~\bibnamefont{Gong}},
  \bibinfo{author}{\bibfnamefont{Y.}~\bibnamefont{Ashida}},
  \bibinfo{author}{\bibfnamefont{K.}~\bibnamefont{Kawabata}},
  \bibinfo{author}{\bibfnamefont{K.}~\bibnamefont{Takasan}},
  \bibinfo{author}{\bibfnamefont{S.}~\bibnamefont{Higashikawa}},
  \bibnamefont{and} \bibinfo{author}{\bibfnamefont{M.}~\bibnamefont{Ueda}},
  \emph{\bibinfo{title}{Topological phases of non-hermitian systems}},
  \bibinfo{journal}{Phys. Rev. X} \textbf{\bibinfo{volume}{8}},
  \bibinfo{pages}{031079} (\bibinfo{year}{2018}).

\bibitem{ashidasinegordon}
\bibinfo{author}{\bibfnamefont{Y.}~\bibnamefont{Ashida}},
  \bibinfo{author}{\bibfnamefont{S.}~\bibnamefont{Furukawa}}, \bibnamefont{and}
  \bibinfo{author}{\bibfnamefont{M.}~\bibnamefont{Ueda}},
  \emph{\bibinfo{title}{Parity-time-symmetric quantum critical phenomena}},
  \bibinfo{journal}{Nat. Commun.} \textbf{\bibinfo{volume}{8}},
  \bibinfo{pages}{15791} (\bibinfo{year}{2017}).

\bibitem{takasu}
\bibinfo{author}{\bibfnamefont{Y.}~\bibnamefont{Takasu}},
  \bibinfo{author}{\bibfnamefont{T.}~\bibnamefont{Yagami}},
  \bibinfo{author}{\bibfnamefont{Y.}~\bibnamefont{Ashida}},
  \bibinfo{author}{\bibfnamefont{R.}~\bibnamefont{Hamazaki}},
  \bibinfo{author}{\bibfnamefont{Y.}~\bibnamefont{Kuno}}, \bibnamefont{and}
  \bibinfo{author}{\bibfnamefont{Y.}~\bibnamefont{Takahashi}},
  \emph{\bibinfo{title}{{PT-symmetric non-Hermitian quantum many-body system
  using ultracold atoms in an optical lattice with controlled dissipation}}},
  \bibinfo{journal}{Progress of Theoretical and Experimental Physics}
  \textbf{\bibinfo{volume}{2020}}(\bibinfo{number}{12}) (\bibinfo{year}{2020}),
  \bibinfo{note}{12A110}.

\bibitem{Peschel2009}
\bibinfo{author}{\bibfnamefont{I.}~\bibnamefont{Peschel}} \bibnamefont{and}
  \bibinfo{author}{\bibfnamefont{V.}~\bibnamefont{Eisler}},
  \emph{\bibinfo{title}{Reduced density matrices and entanglement entropy in
  free lattice models}}, \bibinfo{journal}{Journal of Physics A: Mathematical
  and Theoretical} \textbf{\bibinfo{volume}{42}}(\bibinfo{number}{50}),
  \bibinfo{pages}{504003} (\bibinfo{year}{2009}).

\bibitem{herviou}
\bibinfo{author}{\bibfnamefont{L.}~\bibnamefont{Herviou}},
  \bibinfo{author}{\bibfnamefont{N.}~\bibnamefont{Regnault}}, \bibnamefont{and}
  \bibinfo{author}{\bibfnamefont{J.~H.} \bibnamefont{Bardarson}},
  \emph{\bibinfo{title}{{Entanglement spectrum and symmetries in non-Hermitian
  fermionic non-interacting models}}}, \bibinfo{journal}{SciPost Phys.}
  \textbf{\bibinfo{volume}{7}}, \bibinfo{pages}{69} (\bibinfo{year}{2019}).

\bibitem{chang20}
\bibinfo{author}{\bibfnamefont{P.-Y.} \bibnamefont{Chang}},
  \bibinfo{author}{\bibfnamefont{J.-S.} \bibnamefont{You}},
  \bibinfo{author}{\bibfnamefont{X.}~\bibnamefont{Wen}}, \bibnamefont{and}
  \bibinfo{author}{\bibfnamefont{S.}~\bibnamefont{Ryu}},
  \emph{\bibinfo{title}{Entanglement spectrum and entropy in topological
  non-hermitian systems and nonunitary conformal field theory}},
  \bibinfo{journal}{Phys. Rev. Research} \textbf{\bibinfo{volume}{2}},
  \bibinfo{pages}{033069} (\bibinfo{year}{2020}).

\bibitem{maity}
\bibinfo{author}{\bibfnamefont{S.}~\bibnamefont{Maity}},
  \bibinfo{author}{\bibfnamefont{S.}~\bibnamefont{Bandyopadhyay}},
  \bibinfo{author}{\bibfnamefont{S.}~\bibnamefont{Bhattacharjee}},
  \bibnamefont{and} \bibinfo{author}{\bibfnamefont{A.}~\bibnamefont{Dutta}},
  \emph{\bibinfo{title}{Growth of mutual information in a quenched
  one-dimensional open quantum many-body system}}, \bibinfo{journal}{Phys. Rev.
  B} \textbf{\bibinfo{volume}{101}}, \bibinfo{pages}{180301}
  (\bibinfo{year}{2020}).

\bibitem{calabrese2004}
\bibinfo{author}{\bibfnamefont{P.}~\bibnamefont{Calabrese}} \bibnamefont{and}
  \bibinfo{author}{\bibfnamefont{J.}~\bibnamefont{Cardy}},
  \emph{\bibinfo{title}{Entanglement entropy and quantum field theory}},
  \bibinfo{journal}{Journal of Statistical Mechanics: Theory and Experiment}
  \textbf{\bibinfo{volume}{2004}}(\bibinfo{number}{06}),
  \bibinfo{pages}{P06002} (\bibinfo{year}{2004}).

\bibitem{Pollmann-2009}
\bibinfo{author}{\bibfnamefont{F.}~\bibnamefont{Pollmann}},
  \bibinfo{author}{\bibfnamefont{S.}~\bibnamefont{Mukerjee}},
  \bibinfo{author}{\bibfnamefont{A.~A.~M.} \bibnamefont{Turner}},
  \bibnamefont{and} \bibinfo{author}{\bibfnamefont{J.~E.} \bibnamefont{Moore}},
  \emph{\bibinfo{title}{Theory of finite-entanglement scaling at
  one-dimensional quantum critical points}}, \bibinfo{journal}{Phys. Rev.
  Lett.} \textbf{\bibinfo{volume}{102}}, \bibinfo{eid}{255701}
  (\bibinfo{year}{2009}).

\bibitem{eisert}
\bibinfo{author}{\bibfnamefont{J.}~\bibnamefont{Eisert}},
  \bibinfo{author}{\bibfnamefont{M.}~\bibnamefont{Cramer}}, \bibnamefont{and}
  \bibinfo{author}{\bibfnamefont{M.~B.} \bibnamefont{Plenio}},
  \emph{\bibinfo{title}{\textit{Colloquium} : Area laws for the entanglement
  entropy}}, \bibinfo{journal}{Rev. Mod. Phys.} \textbf{\bibinfo{volume}{82}},
  \bibinfo{pages}{277} (\bibinfo{year}{2010}).

\bibitem{amico}
\bibinfo{author}{\bibfnamefont{L.}~\bibnamefont{Amico}},
  \bibinfo{author}{\bibfnamefont{R.}~\bibnamefont{Fazio}},
  \bibinfo{author}{\bibfnamefont{A.}~\bibnamefont{Osterloh}}, \bibnamefont{and}
  \bibinfo{author}{\bibfnamefont{V.}~\bibnamefont{Vedral}},
  \emph{\bibinfo{title}{Entanglement in many-body systems}},
  \bibinfo{journal}{Rev. Mod. Phys.} \textbf{\bibinfo{volume}{80}},
  \bibinfo{pages}{517} (\bibinfo{year}{2008}).

\bibitem{srednicki}
\bibinfo{author}{\bibfnamefont{M.}~\bibnamefont{Srednicki}},
  \emph{\bibinfo{title}{Entropy and area}}, \bibinfo{journal}{Phys. Rev. Lett.}
  \textbf{\bibinfo{volume}{71}}, \bibinfo{pages}{666} (\bibinfo{year}{1993}).

\bibitem{Bianchini14}
\bibinfo{author}{\bibfnamefont{D.}~\bibnamefont{Bianchini}},
  \bibinfo{author}{\bibfnamefont{O.}~\bibnamefont{Castro-Alvaredo}},
  \bibinfo{author}{\bibfnamefont{B.}~\bibnamefont{Doyon}},
  \bibinfo{author}{\bibfnamefont{E.}~\bibnamefont{Levi}}, \bibnamefont{and}
  \bibinfo{author}{\bibfnamefont{F.}~\bibnamefont{Ravanini}},
  \emph{\bibinfo{title}{Entanglement entropy of non-unitary conformal field
  theory}}, \bibinfo{journal}{Journal of Physics A: Mathematical and
  Theoretical} \textbf{\bibinfo{volume}{48}}(\bibinfo{number}{4}),
  \bibinfo{pages}{04FT01} (\bibinfo{year}{2014}).

\bibitem{Bianchini16}
\bibinfo{author}{\bibfnamefont{D.}~\bibnamefont{Bianchini}} \bibnamefont{and}
  \bibinfo{author}{\bibfnamefont{F.}~\bibnamefont{Ravanini}},
  \emph{\bibinfo{title}{Entanglement entropy from corner transfer matrix in
  forrester{\textendash}baxter non-unitary {RSOS} models}},
  \bibinfo{journal}{Journal of Physics A: Mathematical and Theoretical}
  \textbf{\bibinfo{volume}{49}}(\bibinfo{number}{15}), \bibinfo{pages}{154005}
  (\bibinfo{year}{2016}).

\bibitem{Couvreur}
\bibinfo{author}{\bibfnamefont{R.}~\bibnamefont{Couvreur}},
  \bibinfo{author}{\bibfnamefont{J.~L.} \bibnamefont{Jacobsen}},
  \bibnamefont{and} \bibinfo{author}{\bibfnamefont{H.}~\bibnamefont{Saleur}},
  \emph{\bibinfo{title}{Entanglement in nonunitary quantum critical spin
  chains}}, \bibinfo{journal}{Phys. Rev. Lett.} \textbf{\bibinfo{volume}{119}},
  \bibinfo{pages}{040601} (\bibinfo{year}{2017}).

\bibitem{Dupic}
\bibinfo{author}{\bibfnamefont{T.}~\bibnamefont{Dupic}},
  \bibinfo{author}{\bibfnamefont{B.}~\bibnamefont{Estienne}}, \bibnamefont{and}
  \bibinfo{author}{\bibfnamefont{Y.}~\bibnamefont{Ikhlef}},
  \emph{\bibinfo{title}{{Entanglement entropies of minimal models from
  null-vectors}}}, \bibinfo{journal}{SciPost Phys.}
  \textbf{\bibinfo{volume}{4}}, \bibinfo{pages}{31} (\bibinfo{year}{2018}).

\bibitem{lutheremery}
\bibinfo{author}{\bibfnamefont{A.}~\bibnamefont{Luther}} \bibnamefont{and}
  \bibinfo{author}{\bibfnamefont{V.~J.} \bibnamefont{Emery}},
  \emph{\bibinfo{title}{Backward scattering in the one-dimensional electron
  gas}}, \bibinfo{journal}{Phys. Rev. Lett.} \textbf{\bibinfo{volume}{33}},
  \bibinfo{pages}{589} (\bibinfo{year}{1974}).

\bibitem{Song2020}
\bibinfo{author}{\bibfnamefont{W.}~\bibnamefont{Song}},
  \bibinfo{author}{\bibfnamefont{S.}~\bibnamefont{Gao}},
  \bibinfo{author}{\bibfnamefont{H.}~\bibnamefont{Li}},
  \bibinfo{author}{\bibfnamefont{C.}~\bibnamefont{Chen}},
  \bibinfo{author}{\bibfnamefont{S.}~\bibnamefont{Wu}},
  \bibinfo{author}{\bibfnamefont{S.}~\bibnamefont{Zhu}}, \bibnamefont{and}
  \bibinfo{author}{\bibfnamefont{T.}~\bibnamefont{Li}},
  \emph{\bibinfo{title}{Demonstration of imaginary-mass particles by optical
  simulation in non-hermitian systems}} \bibinfo{note}{ArXiv:2011.08496}.

\bibitem{Lamata2007}
\bibinfo{author}{\bibfnamefont{L.}~\bibnamefont{Lamata}},
  \bibinfo{author}{\bibfnamefont{J.}~\bibnamefont{Le\'on}},
  \bibinfo{author}{\bibfnamefont{T.}~\bibnamefont{Sch\"atz}}, \bibnamefont{and}
  \bibinfo{author}{\bibfnamefont{E.}~\bibnamefont{Solano}},
  \emph{\bibinfo{title}{Dirac equation and quantum relativistic effects in a
  single trapped ion}}, \bibinfo{journal}{Phys. Rev. Lett.}
  \textbf{\bibinfo{volume}{98}}, \bibinfo{pages}{253005}
  (\bibinfo{year}{2007}).

\bibitem{Gerritsma2010}
\bibinfo{author}{\bibfnamefont{R.}~\bibnamefont{Gerritsma}},
  \bibinfo{author}{\bibfnamefont{G.}~\bibnamefont{Kirchmair}},
  \bibinfo{author}{\bibfnamefont{F.}~\bibnamefont{Zähringer}},
  \bibinfo{author}{\bibfnamefont{E.}~\bibnamefont{Solano}},
  \bibinfo{author}{\bibfnamefont{R.}~\bibnamefont{Blatt}}, \bibnamefont{and}
  \bibinfo{author}{\bibfnamefont{C.~F.} \bibnamefont{Roos}},
  \emph{\bibinfo{title}{Quantum simulation of the dirac equation}},
  \bibinfo{journal}{Nature}
  \textbf{\bibinfo{volume}{463}}(\bibinfo{number}{7277}), \bibinfo{pages}{68}
  (\bibinfo{year}{2010}).

\bibitem{Lee2015}
\bibinfo{author}{\bibfnamefont{T.~E.} \bibnamefont{Lee}},
  \bibinfo{author}{\bibfnamefont{U.}~\bibnamefont{Alvarez-Rodriguez}},
  \bibinfo{author}{\bibfnamefont{X.-H.} \bibnamefont{Cheng}},
  \bibinfo{author}{\bibfnamefont{L.}~\bibnamefont{Lamata}}, \bibnamefont{and}
  \bibinfo{author}{\bibfnamefont{E.}~\bibnamefont{Solano}},
  \emph{\bibinfo{title}{Tachyon physics with trapped ions}},
  \bibinfo{journal}{Phys. Rev. A} \textbf{\bibinfo{volume}{92}},
  \bibinfo{pages}{032129} (\bibinfo{year}{2015}).

\bibitem{carmichael}
\bibinfo{author}{\bibfnamefont{H.}~\bibnamefont{Carmichael}},
  \emph{\bibinfo{title}{An Open Systems Approach to Quantum Optics}}
  (\bibinfo{publisher}{Springer-Verlag}, \bibinfo{address}{Berlin},
  \bibinfo{year}{1993}).

\bibitem{PRXQ}
\bibinfo{author}{\bibfnamefont{L.}~\bibnamefont{Xiao}},
  \bibinfo{author}{\bibfnamefont{D.}~\bibnamefont{Qu}},
  \bibinfo{author}{\bibfnamefont{K.}~\bibnamefont{Wang}},
  \bibinfo{author}{\bibfnamefont{H.-W.} \bibnamefont{Li}},
  \bibinfo{author}{\bibfnamefont{J.-Y.} \bibnamefont{Dai}},
  \bibinfo{author}{\bibfnamefont{B.}~\bibnamefont{D\'ora}},
  \bibinfo{author}{\bibfnamefont{M.}~\bibnamefont{Heyl}},
  \bibinfo{author}{\bibfnamefont{R.}~\bibnamefont{Moessner}},
  \bibinfo{author}{\bibfnamefont{W.}~\bibnamefont{Yi}}, \bibnamefont{and}
  \bibinfo{author}{\bibfnamefont{P.}~\bibnamefont{Xue}},
  \emph{\bibinfo{title}{Non-hermitian kibble-zurek mechanism with tunable
  complexity in single-photon interferometry}}, \bibinfo{journal}{PRX Quantum}
  \textbf{\bibinfo{volume}{2}}, \bibinfo{pages}{020313} (\bibinfo{year}{2021}).

\bibitem{bender2006}
\bibinfo{author}{\bibfnamefont{C.~M.} \bibnamefont{Bender}},
  \bibinfo{author}{\bibfnamefont{J.-H.} \bibnamefont{Chen}}, \bibnamefont{and}
  \bibinfo{author}{\bibfnamefont{K.~A.} \bibnamefont{Milton}},
  \emph{\bibinfo{title}{Pt-symmetric versus hermitian formulations of quantum
  mechanics}}, \bibinfo{journal}{Journal of Physics A: Mathematical and
  General} \textbf{\bibinfo{volume}{39}}(\bibinfo{number}{7}),
  \bibinfo{pages}{1657} (\bibinfo{year}{2006}).

\bibitem{ricemele}
\bibinfo{author}{\bibfnamefont{M.~J.} \bibnamefont{Rice}} \bibnamefont{and}
  \bibinfo{author}{\bibfnamefont{E.~J.} \bibnamefont{Mele}},
  \emph{\bibinfo{title}{Elementary excitations of a linearly conjugated
  diatomic polymer}}, \bibinfo{journal}{Phys. Rev. Lett.}
  \textbf{\bibinfo{volume}{49}}, \bibinfo{pages}{1455} (\bibinfo{year}{1982}).

\end{thebibliography}

\pagebreak

\section{Supplementary material for "Correlations at PT-symmetric quantum critical point"}

\setcounter{equation}{0}
\renewcommand{\theequation}{S\arabic{equation}}

\setcounter{figure}{0}
\renewcommand{\thefigure}{S\arabic{figure}}

\section{Calculation of the Green's function of $H$}

The momentum space version of $H$ reads as
\begin{gather}
H_p=\left[\begin{array}{cc}
vp & \Delta\\
0 & -vp
\end{array}\right],
\end{gather}
which has eigenvalues and right eigenvectors as
\begin{subequations}
\begin{gather}
\epsilon_1(p)=vp,\hspace*{3mm}
\Psi_1(p)=\left(\begin{array}{c}
1\\
0\end{array}\right)\\
\epsilon_2(p)=-vp,\hspace*{1mm}
\Psi_2(p)=\frac{1}{\sqrt{\Delta^2+(2vp)^2}}\left(\begin{array}{c}
\Delta\\
-2vp\end{array}\right).
\end{gather}
\label{eigen}
\end{subequations}

In order to contrast our system to those studied in Refs. \cite{Bianchini14,Bianchini16,Dupic}, we note that the right eigenstates of our $H_p^+$ are distinct from those in Eq. \eqref{eigen}, while in the previous works, the ground states of both $H$ and $H^+$ are equal.
In particular, our $H_p^+$ has right eigenstates as
\begin{subequations}
\begin{gather}
\epsilon_1(p)=vp,\hspace*{3mm}
\tilde\Psi_1(p)=\frac{1}{\sqrt{\Delta^2+(2vp)^2}}\left(\begin{array}{c}
2vp\\
\Delta\end{array}\right)\\
\epsilon_2(p)=-vp,\hspace*{3mm}
\tilde\Psi_2(p)=\left(\begin{array}{c}
0\\
1\end{array}\right).
\end{gather}
\end{subequations}

The system is half filled and the ground state of $H_p$ is the minimal energy configuration, therefore it is $\Psi_1(p)$ for $p<0$ and $\Psi_2(p)$ for $p>0$.
Then, the normal Green's function, $G(x)\equiv \langle R^+(x)R(0)\rangle$ is evaluated as
\begin{gather}
G(x)=\sum_{p<0}\Psi_1^+(p)\left[\begin{array}{cc}
1& 0\\
0&0
\end{array}\right]\Psi_1(p)\frac{e^{ipx}}{L}+\nonumber\\
+
\sum_{p>0}\Psi_2^+(p)\left[\begin{array}{cc}
1& 0\\
0&0
\end{array}\right]\Psi_2(p)\frac{e^{ipx}}{L}
\end{gather}
with $L$ the system size. In the thermodynamic limit, summation is transformed to an integral and we recover Eq. (4) in the main text. The calculation of the anomalous Green's function follows similarly as
\begin{gather}
F(x)=\sum_{p>0}\Psi_2^+(p)\left[\begin{array}{cc}
0& 1\\
0&0
\end{array}\right]\Psi_2(p)\frac{e^{ipx}}{L},
\end{gather}
yielding Eq. (6) in the main text. Note that $\Psi_1(p)$ gives vanishing contribution to the anomalous propagator.

\section{Properties of the tight binding model $H^{tb}$}

The tight binding realization of our system is given by
\begin{gather}
H^{tb}=\sum_{n}\frac{J+i\delta(-1)^n}{2}\left(c^+_nc_{n+1}+c^+_{n+1}c_n\right)+\nonumber\\
+(-1)^n g c^+_nc_n.
\end{gather}
For an even number of sites with periodic boundary condition and for an odd number of sites with open boundary condition, the system is PT-symmetric with real spectrum and realizes a 
PT-symmetric quantum critical point at $g=\delta$. 
For the tight binding case, parity transforms $c_j\rightarrow c_{L-j+1}$ while time reversal results in $c_j \rightarrow c_j$ and also $i\rightarrow  -i$.

\begin{figure}[t!]
\centering
\includegraphics[width=7cm]{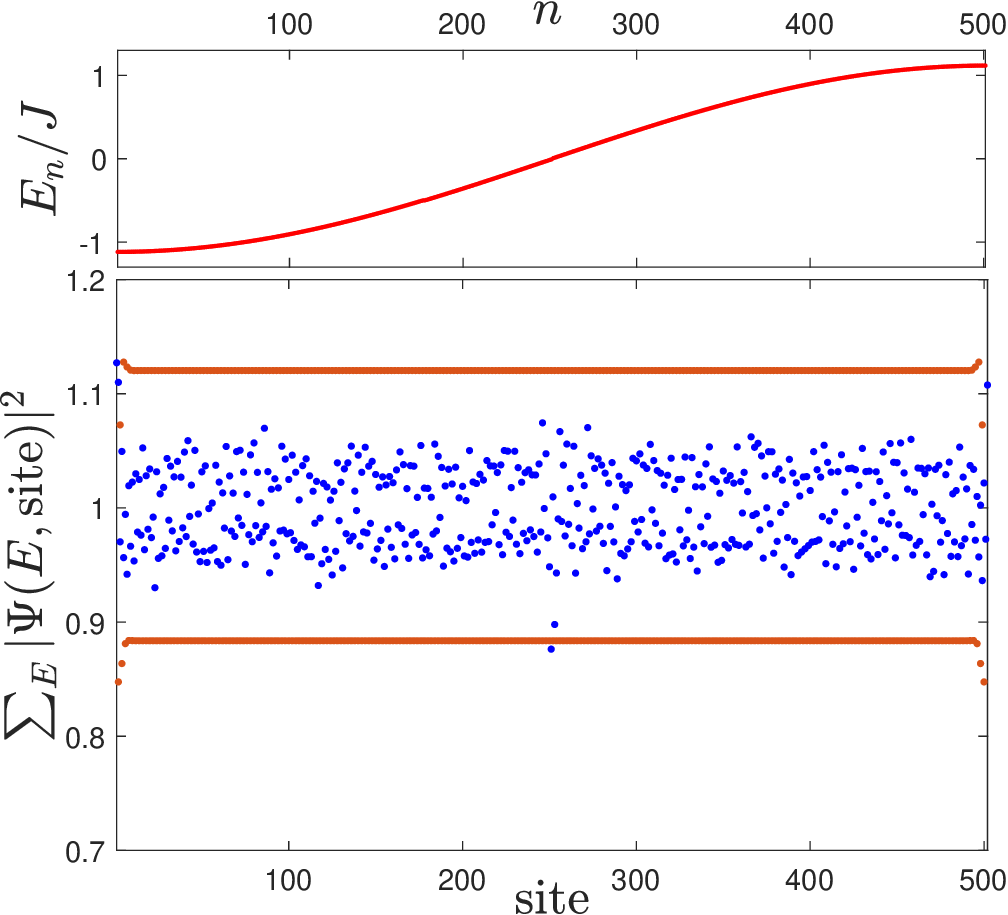}
\caption{(Top panel) Single particle energy eigenvalues $E_n$ of $H^{tb}$ with 502 sites (blue dots) 
and periodic boundary condition and 501 sites with open boundary condition (red dots) with $g=\delta=0.5J$. 
The two data overlap almost perfectly, all eigenvalues are real to numerical precision.
(Bottom panel) Sum of absolute squares of amplitudes per site of all right eigenstates of $H^{tb}$
 with 502 sites (blue dots) and periodic boundary condition and 501 sites with open boundary condition 
(red dots) as in Ref. \cite{Bergholtz2021}. No sign of skin effect is visible.  }
\label{energytb}
\end{figure}

In Fig. \ref{energytb}, the energy eigenvalues are plotted at the critical point with both boundary condition, demonstrating the the presence of critical states.
The lower panel depicts the sum of absolute squares of amplitudes per site of all right eigenstates\cite{Bergholtz2021} of $H^{tb}$, showing no sign of the non-hermitian skin effect.

\begin{figure}[h!]
\centering
\includegraphics[width=8.5cm]{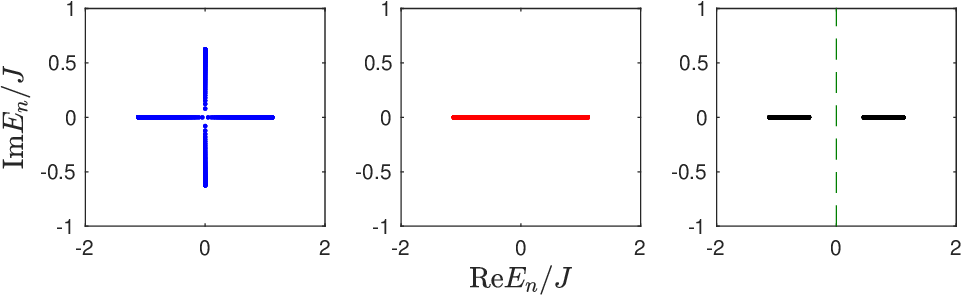}
\caption{Representative evolution of the single particle spectrum is shown across the PT-symmetric critical point for 1002 sites and $g=0.5J$. Left panel: $\delta=0.8J$, PT-symmetry is broken
and there are two 2nd order exceptional points at $E=0$, but separated in momentum space. Upon merging them by approaching $\delta\rightarrow g$, 
 the PT-symmetric critical point is reached (middle panel) with $\delta=0.5J$. By further decreasing $\delta$, PT-symmetry is preserved (right panel, $\delta=0.2$),
and a line gap (green dashed line) appears.}
\label{spec}
\end{figure}

As to the topological properties of this model and its continuum version, the PT-symmetry broken regime ($\delta>g$) 
is gapless, contains two 2nd order exceptional points and is thus expected to be topologically trivial.
On the other hand, the PT-symmetric, gapped phase ($g>\delta$) is adiabatically connected to the Rice-Mele model\cite{ricemele} by e.g.  changing the 
hopping parameter $i\delta$ from imaginary to real. This procedure can be formalized using Theorem 5.1 in Ref. \cite{ashidareview}. This also indicates that this
PT-symmetric gapped phase belongs to the same topological class as the Rice-Mele model. The evolution of the single particle spectrum is visualized in Fig. \ref{spec}

\section{Bosonization of $H$ with the $\Gamma$ term}
We consider a more general version of Eq. (1) in the main text as
\begin{gather}
H=\int dx ~iv \left(R^+(x)\partial_xR(x)-L^+(x)\partial_xL(x)\right)+\nonumber\\
+\Delta R^+(x)L(x)+\Gamma L^+(x)R(x).
\end{gather}
Upon bosonization, this reduces to
\begin{gather}
H=\int \frac{dx}{2\pi} v\left[(\pi\Pi(x))^2+(\partial_x\phi(x))^2\right]+\nonumber\\
+\frac{\Delta}{2\pi\alpha}e^{-i2\phi(x)}+\frac{\Gamma}{2\pi\alpha}e^{i2\phi(x)}=\nonumber\\
=\int \frac{dx}{2\pi} v\left[(\pi\Pi(x))^2+(\partial_x\phi(x))^2\right]+\nonumber\\
+\frac{\Delta+\Gamma}{2\pi\alpha}\cos(2\phi(x))+i\frac{\Gamma-\Delta}{2\pi\alpha}\sin(2\phi(x)).
\end{gather}
Since we use $\Delta>0$ throughout, a finite $\Gamma>0$ enhances the prefactor of the cosine term 
compared to the sine term, which opens up a gap in the spectrum and PT-symmetry remains preserved. 
On the other hand, $\Gamma<0$ increases the absolute value of the prefactor of the sine term compared 
to the cosine, which yields PT-symmetry breaking immediately. These are in accord with the renormalization group analysis of Ref. \cite{ashidasinegordon}.

\end{document}